\documentclass[prd,a4paper,nofootinbib,
showpacs, 
]{revtex4}
\usepackage{graphicx}
\usepackage{amsfonts}
\usepackage{amssymb}
\usepackage{color}
\def\hbar{\hspace{0pt}\raisebox{1pt}{$-$} \hspace{-7pt} h}

\def\5{\overline 5}

\def\chipt{$\chi$PT}
\def\ewchipt{EW$\chi$PT}
\newcommand{\be}{\begin{equation}}
\newcommand{\ee}{\end{equation}}
\newcommand{\bea}{\begin{eqnarray}}
\newcommand{\eea}{\end{eqnarray}}

\begin{document}
\title[]{Asymptotic safety of the gauged $SU(N)$ nonlinear $\sigma$-model}
\date{\today
}
\author{M. Fabbrichesi$^{\ddag}$}
\author{R. Percacci$^{\dag\ddag}$}
\author{A. Tonero$^{\dag\ddag}$}
\author{O. Zanusso$^{\dag\ddag}$}
\affiliation{$^{\ddag}$INFN, Sezione di Trieste} 
\affiliation{$^{\dag}$Scuola Internazionale Superiore di Studi Avanzati\\
via Bonomea 265, 34136 Trieste, Italy}

\begin{abstract}
\noindent
We study the beta functions of the leading, two-derivative
terms of the left-gauged $SU(N)$ nonlinear $\sigma$-model in $d$ dimensions.  
In $d>2$, we find the usual Gaussian ultraviolet fixed point for the
gauge coupling and an attractive non-Gaussian fixed point for the
Goldstone boson coupling. The position of the latter fixed point
controls the chiral expansion, unitarity and the strength of the tree-level Goldstone boson scattering amplitudes.
Attention is paid to the gauge- and scheme-dependence of the results.
\end{abstract}

\maketitle

\section{Motivations} 
\label{sec:motiv}

Any theory where a global symmetry $G$ is spontaneously broken to some subgroup $H$ at some characteristic energy scale $\Lambda$ can be described
at energies $k<\Lambda$ by a Nonlinear $\sigma$--Model (NL$\sigma$M),
a theory describing the dynamics of a set of scalars with values
in the coset space $G/H$ \cite{ccwz}. 
These scalars are the Goldstone bosons.
Because the coset space is (in general) not a linear space,
the physics of the Goldstone bosons is rather different from that
of scalars carrying linear representations of $G$.
The most important phenomenological application of this theory is
chiral perturbation theory (\chipt) \cite{weinberg1,gl}: it describes the dynamics
of the pions, regarded as Goldstone bosons of the flavor symmetry
$SU(N)_L\times SU(N)_R$, which in QCD is broken to the diagonal
subgroup by the quark condensate.
The theory is characterized by a mass scale $f_\pi$ and for
energies $k<f_\pi$, terms with $n$ derivatives give contributions
that are suppressed by factors $(k/4\pi f_\pi)^n$.
So one can usefully expand the action in powers of derivatives.

When such a theory is coupled to gauge fields for the group $G$,
the physical interpretation changes completely. The Goldstone
bosons are acted upon transitively by the gauge group, which means
that any field configuration can be transformed into any other
field configuration by a gauge transformation.
So, in a sense, they are now gauge degrees of freedom.
It is then possible to fix the gauge in such a way that the Goldstone
bosons disappear completely from the spectrum. 
In this ``unitary'' gauge no residual gauge freedom is left, so the 
spectrum of the theory consists just of massive gauge fields, the masses
originating from the covariant kinetic term of the Goldstone bosons.
This is the essence of the Higgs phenomenon, but in this variant
where the scalars carry a nonlinear realization of $G$,
there is no physical Higgs field left over.

The most important phenomenological application of this idea is
Electroweak Chiral Perturbation Theory (\ewchipt) \cite{ab}.
It is similar to \chipt, except that the ``pions'' are identified with 
the angular degrees of freedom of the Standard Model (SM) Higgs field, 
and are coupled to the electroweak gauge fields.
The pion decay constant is identified with the Higgs VEV $\upsilon$.
The target space is $(SU(2)_L\times U(1)_Y)/U(1)_Q\sim SU(2)$,
just as in the simplest \chipt.
At tree level one can view this as the SM Higgs sector in the limit when 
the quartic coupling $\lambda\to\infty$ at fixed VEV $\upsilon$,
so that the mass of the Higgs field goes to infinity.
Thus \ewchipt\ can be seen as an approximation to the SM
when the energy is sufficiently low that the Higgs degree of freedom
cannot be excited.
This is actually the domain that has been experimentally studied so far.

This model is perfectly adequate to give mass to the gauge bosons.
The main reason why the SM uses a linearly transforming Higgs field,
rather than a nonlinear one,
is that the NL$\sigma$M is perturbatively nonrenormalizable.
In practice, when one computes scattering amplitudes, more and
more terms are needed to achieve a predetermined precision,
as the energy is increased. 
Thus \ewchipt\ becomes less and less useful and eventually the 
perturbative procedure breaks down for momenta of order $4\pi\upsilon$,
where terms with any number of derivatives are equally important.

One might also worry about the experimental fact that
a very heavy Higgs particle is disfavored by precision electroweak data.
If the NL$\sigma$M was really equivalent to the $\lambda\to\infty$ 
limit of the Higgs model, then this would be enough to essentially rule it out.
However, at the quantum level, the two models are not equivalent:
since the coupling grows together with the mass, the decoupling theorem
fails \cite{herrero,chan}.
The issue of the compatibility of the gauged NL$\sigma$M with precision
electroweak data has to be analyzed separately \cite{bagger}.

For these reasons, is important to have a good understanding of the UV behavior of this model.
Here we would like to explore the possibility that the gauged NL$\sigma$M
could be asymptotically safe, which means that its Renormalization Group (RG) 
flow has a fixed point (FP) with a finite number of attractive directions.
As discussed in \cite{weinberg2}, this would make the theory
UV complete and predictive.
So far asymptotic safety has been thought of mainly as a way of
constructing a consistent quantum field theory of gravity \cite{gravity}, 
but for this idea to give a physically viable theory it is necessary 
that also all the other interactions should behave in this way.
There are then two possibilities: one is that asymptotic safety
of all interactions is closely related to gravitational effects
and only manifests itself at the Planck scale \cite{perini}; 
the other that each interaction independently becomes asymptotically safe 
at its own characteristic scale.
We will find here some evidence for the latter behavior.
At least, our findings suggest that the range of validity of 
\ewchipt\ could be extended to higher energies than normally thought, 
in which case one might not see a fundamental Higgs field at the LHC at all.

Asymptotic safety could also manifest itself with a linearly realized Higgs field, 
in the presence of Yukawa interactions \cite{gies}.
In both cases the evidence is not conclusive, but the results indicate 
that this line of thought deserves to be pursued further.
In forthcoming publications we will discuss the more realistic case
of $SU(2)\times U(1)$ gauging, the effect of fermions coupled to the gauged NL$\sigma$M and the compatibility of this model with electroweak precision tests \cite{us}.
In this paper we will not try to derive any phenomenological consequence, 
but merely consider the theoretical problem of a chiral NL$\sigma$M with values
in $SU(N)$ coupled to $SU(N)_L$ gauge fields.

\section{$SU(N)$ gauged nonlinear sigma model} 
\label{sec:nlsm}

The NL$\sigma$M describes the dynamics of  
a map $\varphi$ from spacetime, a $d$-dimensional manifold $\mathcal{M}$,
to a $D $-dimensional target manifold $\mathcal{N}$. 
Given a coordinate system  $\{x^\mu\}$ on $\mathcal{M}$ and $\{y^\alpha\}$
on $\mathcal{N}$, one can describe the map $\varphi$ by $D$ scalar fields  
$\varphi^\alpha(x)$. 
Physics must be independent of the choice of coordinates, 
and this can be achieved by using standard methods of differential geometry. 
We will restrict ourselves to the case $\mathcal{N}=SU(N)$,
endowed with a left- and right-invariant metric $h_{\alpha\beta}$.
In order to describe this geometry we begin by choosing matrix generators 
$\{T_i\}$ in the fundamental representation satisfying 
$[T_i,T_j]=f_{ij}{}^k T_k$
where $f_{ij}{}^k$ are the structure constants.
The $Ad$-invariant Cartan-Killing form is
$B_{ij}=Tr(Ad(T_i)Ad(T_j))=f_{i\ell}{}^k f_{jk}{}^\ell=-N\delta_{ij}$, 
whereas in the fundamental representation $Tr(T_iT_j)=(1/2)\delta_{ij}$.
We choose to work with the inner product in the Lie algebra 
$-(1/N)B_{ij}=\delta_{ij}$.
The $Ad$-invariance of this inner product implies that
$f_{ijk}=f_{ij}{}^\ell\delta_{\ell k}$ is totally antisymmetric.

Under the identification of the Lie algebra with the tangent space
to the group at the identity, to each abstract generator $T_i$ 
there corresponds a left-invariant vectorfield $L_i^\alpha$
and a right-invariant vectorfield $R_i^\alpha$,
coinciding with $T_i$ in the identity.
They satisfy the commutation relations
\be
[L_i,L_j]=f_{ij}{}^k L_k\ ;\qquad
[R_i,R_j]=-f_{ij}{}^k R_k\ .
\ee
They form fields of bases on the group.
We will often use also the dual bases $L^i_\alpha$ and $R^i_\alpha$:
\be
L^i_\alpha L_j^\alpha=\delta^i_j\ ;\qquad
R^i_\alpha R_j^\alpha=\delta^i_j\ ;\qquad
R^i_\alpha L_j^\alpha=Ad(U)^i{}_j\ ,\qquad
\ee
where $U$ denotes the $n\times n$ matrix corresponding to the
group element with coordinate $\varphi$.
The dual bases are the components of the Maurer-Cartan forms:
$L^i T_i=U^{-1}dU$, $R^i T_i=dU U^{-1}$.

The metric $h_{\alpha\beta}$ on the group is defined as the unique left- and right-invariant
metric that coincides with the inner product in the Lie algebra:
$\delta_{ij}=h(\mathbf{1})(R_i,R_j)=h(\mathbf{1})(L_i,L_j)$.
Thus, the vectorfields $R_i$ and $L_i$ are Killing vectors,
generating $SU(N)_L$ and $SU(N)_R$ respectively, 
and they are also orthonormal fields of frames on the group:
\be
\label{orthK}
h_{\alpha\beta}=R^i_\alpha R^j_\beta \delta_{ij}=L^i_\alpha L^j_\beta \delta_{ij}\ .
\ee

We will consider the case when only $SU(N)_L$ is gauged,
and call $A_\mu$ the corresponding gauge field.
The covariant derivative and the gauge field strength are defined to be:
\be
D_\mu \varphi^\alpha=\partial_\mu\varphi^\alpha+A_\mu^i R_i^\alpha(\varphi)\qquad\qquad F^i_{\mu\nu}=\partial_\mu A_\nu^i-\partial_\nu A_\mu^i+f_{jl}{}^i A_\mu^jA_\nu^l\,.
\ee

Restricting our attention to terms containing two derivatives of the fields,
the Euclidean action of this gauged NL$\sigma$M, in $d$ dimensions, reads 
\be
\label{gnlaction}
S_{\rm } = \frac{1}{2f^2}  \int d^dx\, 
h_{\alpha\beta}D_\mu\varphi^\alpha D^\mu\varphi^\beta
+ \frac{1}{4g^2}\int d^dx\, F^i_{\mu\nu}F_i^{\mu\nu},
\ee
where $f$ and $g$ are couplings.
The action (\ref{gnlaction}) is invariant under local $SU(N)_L$ infinitesimal transformation
\be
\label{transf}
\delta_\epsilon\varphi^\alpha=-\epsilon_L^i R_i^\alpha(\varphi)\qquad\qquad
\delta_\epsilon A_{\mu}^i=\partial_\mu\epsilon_L^i+f_{j\ell}{}^i A_{\mu}^j\epsilon_L^\ell\,.
\ee

To evaluate the beta functions we expand around (nonconstant) background fields
$\bar A_\mu^i$ and $\bar\varphi^\alpha$.
The gauge field is expanded as 
$A_\mu^i(x)=\bar A_\mu^i(x)+a_\mu^i(x)$
whereas for the NL$\sigma$M the fluctuation is best described by its 
normal coordinates $\xi^\alpha(x)$ centered at $\bar\varphi^\alpha(x)$:
$\mathrm{Exp}_{\bar\varphi(x)}(\xi(x))=\varphi(x)$ \cite{BG}.
This relation can be expanded as
\be
\label{exponential}
\varphi^\alpha=\bar\varphi^\alpha+\xi^\alpha
-\frac{1}{2}\Gamma_\beta{}^\alpha{}_\gamma \xi^\beta\xi^\gamma+\ldots\ ,
\ee
where $\Gamma_\beta{}^\alpha{}_\gamma$ are the Christoffel symbols of the
metric $h_{\alpha\beta}$ evaluated at $\bar\varphi(x)$.
The background field expansions for the geometric objects entering in (\ref{gnlaction}) are given by \cite{BG}:
\begin{eqnarray}
h_{\alpha\beta}(\varphi)&=& h_{\alpha\beta}(\bar \varphi) -\frac{1}{3} R_{\alpha\epsilon\beta\eta}\xi^\epsilon\xi^\eta+\cdots\nonumber\\
\partial_\mu\varphi^\alpha&=& \partial_\mu\bar{\varphi}^\alpha +\nabla_\mu\xi^\alpha-\frac{1}{3}\partial_\mu\bar{\varphi}^\gamma
R_{\gamma\epsilon}\,\!^\alpha\,\!_\eta \xi^\epsilon\xi^\eta  + \cdots \nonumber\\
R^\alpha_i(\varphi)&=& R^\alpha_i(\bar \varphi) + \xi^\epsilon\nabla_\epsilon R^\alpha_i - \frac{1}{3}
R^\alpha\,\!_{\epsilon\gamma\eta}R^\gamma_i\xi^\epsilon\xi^\eta + \cdots \nonumber\\
D_\mu\varphi^\alpha&=& D_\mu\bar{\varphi}^\alpha 
+\nabla_\mu\xi^\alpha+A_\mu^i\nabla_\epsilon R_i^\alpha\xi^\epsilon
-\frac{1}{3}D_\mu\bar{\varphi}^\gamma
R_{\gamma\epsilon}\,\!^\alpha\,\!_\eta \xi^\epsilon\xi^\eta  
+ \cdots 
\end{eqnarray}
where $\nabla$ is the Riemannian covariant derivative of the metric $h$
and all tensors on the r.h.s. are evaluated at $\bar\varphi$.
The background field expansion for the gauge field strength tensor is given by
\be
F_{\mu\nu}^i=\bar F_{\mu\nu}^i+\bar D_\mu a_\nu^i-\bar D_\nu a_\mu^i + f_{j\ell}{}^i a_\mu^ja_\nu^\ell\, ,
\ee
where $\bar D_\mu a_\nu^i=\partial_\mu a_\nu^i+f_{j\ell}{}^i\bar A_\mu^ja_\nu^\ell$.

To the action we add a background gauge fixing term:
\be
\label{gaugefixing}
S_{gf}=\frac{1}{2\alpha g^2}\int d^dx\, \delta_{ij}\chi^i\chi^j
\qquad\mbox{with}\qquad
\chi^i=\bar D^\mu a^i_\mu + \beta \frac{g^2}{f^2}  R^i_{\alpha}\xi^\alpha\ .
\ee
where $\alpha$ and $\beta$ are gauge fixing parameters.
The case $\alpha=\beta$ is a generalization of $\alpha$-gauge fixing 
(what is usually known as $R_\xi$-gauge) to the background field method. 
Moreover, for $\alpha=\beta=1$ we have the generalization of the 't Hooft-Feynman
gauge fixing.

In order to obtain the Faddeev-Popov operator we need the gauge variations
of the fields, keeping the background fixed. 
In (\ref{transf}) we have given the transformation of the
coordinates $\varphi^\alpha$. The variation of the corresponding normal
coordinates $\xi^\alpha$ can be worked out using (\ref{exponential}):
$\delta\varphi^\alpha=\delta\xi^\alpha
+\Gamma_\beta{}^\alpha{}_\gamma\xi^\beta\delta\xi^\gamma+\ldots$
and inverting this series.
For our purposes only the first term matters:
\be
\delta_{\epsilon_L} \chi^i=
\bar D_\mu \bar D^\mu \epsilon_L^i-\beta\frac{g^2}{f^2} \epsilon_L^i+\ldots\ ,
\ee 
where the dots stand for terms containing $\xi$.
In this way we obtain the ghost action $S_{gh}=S_{ghF}+S_{ghI}$,
where 
\be
S_{ghF}=\int d^dx\, \bar{c}^i \left(-\bar D^2 + \beta \frac{g^2}{f^2}\right)\delta_i^j\, c_j
\ee
is the free ghost action and $S_{ghI}$ are interaction terms.

The action (\ref{gnlaction}) can be expanded in functional Taylor series around the backgrounds:
\be
S(\varphi,A)=
S(\bar\varphi,\bar A)
+S^{(1)}(\bar\varphi,\bar A,;\xi,a)
+S^{(2)}(\bar\varphi,\bar A,;\xi,a)+\ldots
\ee 
where $S^{(n)}$ is of order $n$ in the fluctuations.
The second order piece is
\begin{eqnarray}\label{BFaction}
S^{(2)}&=&
\frac{1}{2f^2} \int d^dx\,\xi^\alpha \left(-\bar D^2h_{\alpha\beta} -D_\mu\bar{\varphi}^\epsilon D^\mu \bar{\varphi}^\eta R_{\epsilon\alpha\eta\beta} \right)\xi^\beta 
+\frac{1}{f^2} \int d^dx\, a_\mu^i \left( h_{\alpha\gamma} D^\mu\bar{\varphi}^\alpha\nabla_\beta R^\gamma_i 
+h_{\alpha\beta} R^\alpha_i \bar D^\mu \right) \xi^\beta 
\nonumber\\
&+& \frac{1}{2g^2} \int d^dx\, a_\mu^i \left(-\bar D^2\delta_{ij}\delta^{\mu\nu}+\bar D^\nu \bar D^\mu\delta_{ij} 
 +\frac{g^2}{f^2} h_{\alpha\beta}R^\alpha_i R^\beta_j \delta^{\mu\nu} 
 + \bar{F}^{\ell\mu\nu}f_{\ell ij} \right) a_\nu^j \, ,
\end{eqnarray}
where $\bar D_\mu\xi^\alpha=\nabla_\mu\xi^\alpha + \bar{A}_\mu^i\nabla_\beta R_i^\alpha \xi^\beta\,$ .

In the following we drop all bars from background quantities,
since no confusion should arise.
In the second integral we can perform an integration by parts to remove
the derivative on $\xi$, and use the Killing property 
$\nabla_\alpha R^i_\beta=-\nabla_\beta R^i_\alpha$
to rewrite the mixed terms as
$$
\frac{1}{f^2} \int d^dx\, 
\left(2a_\mu^i D^\mu {\varphi}^\alpha\nabla_\beta R^i_\alpha-R^i_\beta D^\mu a_\mu^i\right)
\xi^\beta 
$$
Now we add the gauge fixing and the free ghost action to obtain the
complete gauge fixed quadratic action:
\begin{eqnarray}
\label{gfaction}
S^{(2)}&=&
\frac{1}{2f^2} \int d^dx\,
\xi^\alpha 
\left(-D^2h_{\alpha\beta} -D_\mu {\varphi}^\epsilon D^\mu {\varphi}^\eta R_{\epsilon\alpha\eta\beta}
+ \frac{\beta^2}{\alpha}\frac{ g^2}{f^2} h_{\alpha\beta}
 \right)\xi^\beta 
\nonumber\\
&+& \frac{1}{2g^2} \int d^dx\, 
a_\mu^i \left(-D^2\delta_{ij}\delta^{\mu\nu} 
+\left(1-\frac{1}{\alpha}\right)\delta_{ij}D^\mu D^\nu
+\frac{g^2}{f^2}\delta_{ij} \delta^{\mu\nu} - 2 F^{\ell\mu\nu}f_{i\ell j}\right) a_\nu^j \nonumber\\
&+& 2\frac{1}{f^2} \int d^dx\, a^{\mu i} D_\mu {\varphi}^\alpha h_{\alpha\gamma}
\nabla_\beta R_i^\gamma\xi^\beta 
+\frac{1}{f^2}\left(\frac{\beta}{\alpha}-1\right)\int d^dx\,D^\mu a_\mu^i\delta_{ij}R^j_\beta\xi^\beta
+ S_{ghF}\, .
\end{eqnarray}

From here onwards we will set $\beta=\alpha$ in order to
get rid of the second mixed term.

At this point it is convenient to define $\xi^i=R^i_\alpha \xi^\alpha $, 
and to  introduce a $D(1+d)$ component bosonic field $\theta^T=(\xi^i,a_\mu^i)$ 
in order to write (\ref{gfaction}) in a more compact form
\begin{eqnarray}
S^{(2)}&=&
\frac{1}{2} \int d^dx\, \theta^T (\mathbb{Q-E})\, \theta 
+ \int d^dx\, \bar{c}^i \left(-D_c^2 + \alpha\frac{g^2}{f^2}\right)\delta^j_i\, c_j\, .
\end{eqnarray}
The differential operator $\mathbb{Q}$ and the block-matrix $\mathbb{E}$ are
\be
\label{gfaction2}
\mathbb{Q} = \left(
\begin{array}{cc}
\frac{1}{f^2}(-D_\xi^2+\alpha \frac{ g^2}{f^2}) & 0 \\
0 & \frac{1}{g^2}\left[(-D_a^2+\frac{ g^2}{f^2})\delta^{\mu\nu}
+\left(1-\frac{1}{\alpha}\right)D^\mu D^\nu\right]
\\
\end{array} \right)\, 
\qquad,\qquad
\mathbb{E} = \left(
\begin{array}{cc}
\frac{1}{f^2} {M}_{ij} & \frac{1}{f^2} {B}^\mu_{ij}\\
\frac{1}{f^2} {B}^{T\mu}_{ij} & \frac{2}{g^2}F^{\mu\nu}_{ij} \\
\end{array} \right)
\ee
where
\be
\label{emme}
M_{ij}= R_i^\alpha R_j^\beta 
D_\mu {\varphi}^\epsilon D^\mu  {\varphi}^\eta R_{\epsilon\alpha\eta\beta}
\ ;\qquad
F_{ij}^{\mu\nu}= {F}^{\ell\mu\nu}f_{i \ell j}
\ ;\qquad
B_{ij}^\mu=- 2D^\mu {\varphi}^\beta R_i^\alpha\nabla_\alpha R_{\beta j} .
\ee


\section{Beta functions}


We compute the beta functions of the theory 
using functional renormalization group methods.
We start from the ``effective average action'' 
$\Gamma_k(\varphi,A;\xi,a,\bar c,c)$,
depending on the background fields $\varphi$ and $A$,
and on the ``classical fields'' $\xi$, $a$, $\bar c$, $c$
(the variables that are Legendre conjugated to the sources coupling linearly
to the quantum field by the same name; we trust this notational abuse
will not be cause of misunderstandings).
The effective average action
is defined exactly as the usual background field effective action except for 
an infrared modification of the propagators.
In the present context, this modification is specified by adding
to the inverse propagators $\frac{\delta^2\Gamma_k}{\delta\theta\delta\theta}$
and $\frac{\delta^2\Gamma_k}{\delta\bar{c}\delta c}$
some kernels $\mathcal{R}^\theta_k$ and $\mathcal{R}^{c}_k$
which go to zero for momenta greater than $k$.
This functional obeys a functional differential equation \cite{Wetterich}
which in the present context reads
\be
\label{betafunctional}
\frac{d\Gamma_k}{dt} = \frac{1}{2}\mathrm{Tr}\left(\frac{\delta^2\Gamma_k}{\delta\theta\delta\theta}+\mathcal{R}^\theta_k\right)^{-1}
\frac{d{\mathcal{R}}^\theta_k}{dt}
-\mathrm{Tr}\left(\frac{\delta^2\Gamma_k}{\delta\bar{c}\delta c}+\mathcal{R}^{c}_k\right)^{-1}\frac{d{\mathcal{R}}^{c}_k}{dt}\, ,
\ee
where $t=\mathrm{log}(k/k_0)$.
This functional  equation is  exact. There is no reference to a bare action and there is no need to introduce an UV regulator, on account of the fact that the properties of the cutoff kernels $\mathcal{R}^\theta_k$ and $\mathcal{R}^c_k$ ensure that the r.h.s. of (\ref{betafunctional}) is UV finite.
For earlier applications of this equation to gauge theories see \cite{gerge}.

In order to extract the beta functions of the theory
we assume that the functional $\Gamma_k(\varphi,A;0,0,\bar c,c)$ 
has the form of the functional $S$ introduced in (\ref{gnlaction}), 
with the bare couplings $f$ and $g$ replaced by renormalized coupling that depend on $k$,
plus the gauge fixing action (\ref{gaugefixing}),
with the gauge parameters assumed to be fixed.
Inserting this ansatz for $\Gamma_k$ in (\ref{betafunctional})
will yield the beta functions of $g$ and $f$.
Notice that we will be reading off the beta functions from the background
field monomials.
This truncation of $\Gamma_k$ is strictly speaking not consistent, because
the beta functions of the couplings that are being neglected are not zero.
In any case this method reproduces the results of perturbation theory when 
suitable approximations are made and it is ``nonperturbative'' 
in the sense that it does not rely on the couplings being small.
Improved results can be obtained by keeping more terms in the average effective action.
This method has been applied to the NL$\sigma$M with two and four derivatives 
in \cite{codello} and \cite{zanusso} respectively.
The novelty here is the presence of the gauge field.

There is a lot of freedom in the choice of the cutoff kernels.
Generally, one chooses them in such a way as to make the calculations simpler,
but it is also interesting to examine the dependence of the results on such choices.
We will refer to this as ``scheme dependence'', because in the context of perturbation 
theory it is closely related to the dependence of results on the renormalization scheme.
Results that have a direct physical significance should be scheme independent.
We will calculate the beta functions in two different cases.
The first calculation uses the 't Hooft-Feynman gauge $\alpha=1$
and is valid in any dimension.
The second calculation is in an arbitrary $\alpha$-gauge but is
restricted to four dimensions.
It will be convenient to adopt slightly different schemes in the two cases.
We will then compare the results of the two calculations in four dimensions.

\subsection{Arbitrary dimension, 't Hooft-Feynman gauge}

In this subsection we choose $\alpha=1$.
The $a$-$a$ terms in (\ref{gfaction}) or (\ref{gfaction2})
then become a minimal second order operator (the highest order part is a Laplacian)
and this simplifies the calculation significantly.
We choose the cutoff kernels to be functions of the background covariant Laplacians, of the form
\be
\mathcal{R}^\theta_{k} = \left(
\begin{array}{cc}
\frac{1}{f^2} R_k(-D_\xi^2) & 0 \\
0 & \frac{1}{g^2}R_k(-D_a^2) \\
\end{array} \right)\qquad,\qquad
\mathcal{R}^{c}_{k }= R_k(-D_c^2)\, .
\ee
In the terminology of \cite{cpr2} this is called a ``type I'' cutoff.
For the cutoff profile function $R_k$ we choose the ``optimized'' form
$R_k(z) = (k^2-z)\theta (k^2-z)$ \cite{optimized}, which ensures that the
integrations over momenta are finite and explicitly calculable.
Being constructed with the background Laplacians,
this cutoff prescription preserves the background gauge invariance.
The $t$-derivative of this cutoff is
\be
\frac{d{\mathcal{R}}^\theta_{k}}{dt} = \left(
\begin{array}{cc}
\frac{1}{f^2} \left[\partial_t{R}_k(-D_\xi^2)+\eta_\xi R_k(-D_\xi^2)\right]& 0 \\
0 & \frac{1}{g^2}\left[\partial_t{R}_k(-D_a^2)+\eta_a R_k(-D_a^2)\right] \\
\end{array} \right)\, ,
\ee
where $\eta_\xi=-2\partial_t\log f$ and $\eta_a=-2\partial_t\log g$.
The other parts of equation (\ref{betafunctional}) are
\be
\frac{\delta^2\Gamma_k}{\delta\theta\delta\theta}+\mathcal{R}^\theta_k
\equiv \mathbb{P}_k-\mathbb{E}\ ;\qquad
\mathbb{P}_k
=\mathbb{Q}+{\mathcal{R}}^\theta_{k}
=\left(
\begin{array}{cc}
\frac{1}{f^2}  (P_k(-D_\xi^2)+\frac{g^2}{f^2}  ) & 0 \\
0 & \frac{1}{g^2}(P_k(-D_a^2)+\frac{g^2}{f^2} ) \\
\end{array} \right)
\ee
and
\be
\frac{\delta^2\Gamma_k}{\delta\bar{c}\delta c}+\mathcal{R}^{c}_k
=P_k(-D_c^2)+\frac{g^2}{f^2} 
\ee
where we defined the function $P_k(z)=z+R_k(z)$, 
which is equal to
$k^2\theta(k^2-z)+z\theta(z-k^2)$ for the optimized cutoff.

By expanding the first term of (\ref{betafunctional}) in powers of $(\mathbb{P}_k)^{-1}\mathbb{E}$, we therefore have
\be
\label{expansion1}
\frac{d{\Gamma}_k}{dt} = 
\frac{1}{2}\mathrm{Tr}\left(\mathbb{P}_k^{-1}
+\mathbb{P}_k^{-1}\mathbb{E}\mathbb{P}_k^{-1}
+\mathbb{P}_k^{-1}\mathbb{E}\mathbb{P}_k^{-1}\mathbb{E}\mathbb{P}_k^{-1}
+\ldots\right)
\frac{d{\mathcal{R}}^\theta_k}{dt} 
-\mathrm{Tr}\left(P_k(-D^2_c)+\frac{g^2}{f^2}\right)^{-1}\frac{d{\mathcal{R}}^{c}_k}{dt}\, . \label{equazione}
\ee
Note that in so doing we keep the entire dependence on $g^2/f^2$ but we expand
in powers of $\mathbb{E}$, which depends on the background fields.
These traces can be evaluated using heat kernel methods,
described e.g. in Appendix A of \cite{cpr2}.

Summing all contributions and reading off the coefficients of
$(1/4)\int F_{\mu\nu}^i F^{i\mu\nu}$ and
$(1/2)\int D_\mu\varphi^\alpha D^\mu\varphi^\beta h_{\alpha\beta}$
we obtain the beta functions for $1/g^2$ and ${1}/{f^2} $: 
\bea
\label{betag1}
\frac{d}{dt}\frac{1}{g^2}&=&
-\frac{1}{(4\pi)^{d/2}}\frac{N}{3}
\frac{1}{\Gamma\left(\frac{d}{2}-1\right)}
\frac{k^{d-4}}{1+\frac{\tilde g^2}{\tilde f^2}}
\left[
\frac{1}{4}+d-2+\frac{\eta_\xi/4+d\eta_a}{d-2} 
-\frac{192}{d(d-2)}
\frac{1}{(1+ \frac{\tilde g^2}{\tilde f^2})^2}\left(1+\frac{\eta_a}{d+2}\right)\right]\, , \\
\label{betazeta1}
\frac{d}{dt}\frac{1}{f^2}  &=&
\frac{1}{(4\pi)^{d/2}}\frac{N}{2}
\frac{1}{\Gamma\left(\frac{d}{2}+1\right)}
\frac{k^{d-2}}{(1+\frac{\tilde g^2}{\tilde f^2} )^2}
\left[
1+\frac{\eta_\xi}{d+2} +\frac{4\tilde g^2/\tilde f^2 }{1+ \frac{\tilde g^2}{\tilde f^2}}
\left(2+\frac{\eta_\xi+\eta_a}{d+2}\right)\right]\,. \label{RG1}
\eea

A few comments are in order at this point.
As they stand, these are not yet explicit beta functions, because the right hand sides
contain the beta functions themselves inside the factors of $\eta_\xi$ and $\eta_a$.
Thus, these can be regarded as algebraic equations for the beta functions.
We do not give the explicit expressions of the beta functions here because they are somewhat 
unwieldy and can easily be obtained by solving the above equations.
We observe that omitting the terms containing $\eta_\xi$ and $\eta_a$
in the right hand sides gives the one loop beta functions.

These equations give the beta functions of the (generally) dimensionful couplings.
The corresponding beta functions of the dimensionless combinations
$\tilde f^2= f^2 k^{d-2}$ and $\tilde g^2=g^2 k^{d-4}$ can be obtained by simple algebra.
Note that on the right hand side the dimensions are carried just by the explicit
powers of $k$, all the rest is dimensionless.

As we mentioned earlier, the only approximation made in this
calculation consists in neglecting higher derivative terms.
We know that this is a good approximation at sufficiently low energy and
we are implicitly assuming that it remains a reasonably good approximation
also at higher energy.
Provided this important assumption is true, these beta functions are valid at all energy scales:
having used a mass-dependent renormalization we get automatically the
effect of thresholds, which are represented by the factors $1/(1+ g^2/\tilde f^2)$
(note that $g^2/f^2$ has dimensions of mass squared in any dimension).
For $k^2\gg g^2/f^2$ these factors become equal to one,
whereas for $k^2\ll  g^2/f^2$ the denominators become large and suppress the running,
reflecting the decoupling of the corresponding massive field modes.

Finally we observe that (\ref{betag1}) has an apparent pole at $d=2$,
which is actually cancelled by the pole of the function $\Gamma(d/2-1)$
in the denominator.

\subsection{Four dimensions, generic $R_\xi$-gauge}

We now consider a general $R_\xi$ gauge, where the parameter $\xi$ is now called $\alpha$
in order not to generate confusion with the Goldstone modes.
In this case the operator acting on the gauge fluctuations is nonminimal
(meaning that the highest order terms are not simply a Laplacian).
Due to the increased complication, from here on we will restrict ourselves to four dimensions.
A standard way of dealing with nonminimal operators is to
decompose the field they act on in irreducible components, in the present case the
longitudinal and transverse parts of $a$.
The resulting operators acting on the irreducible subspaces are typically of Laplace type.
We thus define operators ${\cal D}_L$ and ${\cal D}_T$ by
\be
{\cal D}_{T\mu\nu}=-D^2\delta_{\mu\nu}-2F_{\mu\nu}\ ;\qquad
{\cal D}_{L\mu\nu}=-D_\mu D_\nu\ ,
\ee
where it is understood that $F$ acts to the fields on its right in the adjoint representation,
as in equation (\ref{emme}).
Assuming that the background gauge field is covariantly constant,
one easily proves that the following operators are projectors:
\be
\mathbf{P}_L={\cal D}_T^{-1}{\cal D}_L\ ;\qquad
\mathbf{P}_T=\mathbf{1}-\mathbf{P}_L\ .
\ee
We can then decompose $a_\mu=t_\mu+D_\mu\psi$, where 
$t_\mu=\mathbf{P}_T a_\mu$, $D_\mu t^\mu=0$
and $\psi=-D_\rho{\cal D}_{T\rho\sigma}^{-1}a_\sigma$.
We then introduce the cutoff separately in the transverse and longitudinal subspaces as follows:
\be
\mathcal{R}_{k}^\theta = \left(
\begin{array}{cc}
\frac{1}{f^2} R_k(-D^2-M) & 0 \\
0 & \frac{1}{g^2}\left[ R_k({\cal D}_T)\mathbf{P}_T 
+\frac{1}{\alpha} R_k({\cal D}_T)\mathbf{P}_L\right] \\
\end{array} \right)\, .
\ee
Note that the cutoff is now a function of the kinetic operator
acting in each irreducible subspace, including
the background-dependent terms $M$ and $F$, but not the mass-like term $g^2/f^2$.
Following the terminology of \cite{cpr2}, we call this a ``type II cutoff''.
For the ghosts we use the same cutoff as before. 

The modified bosonic inverse propagator is now
\be
\frac{\delta^2S}{\delta\theta\delta\theta}+\mathcal{R}_k^\theta=
\left(
\begin{array}{cc}
\frac{1}{f^2} (P_k(-D^2-M)+\alpha\frac{g^2}{f^2} )& B \\
B^T & \frac{1}{g^2}\left[(P_k({\cal D}_T)+\frac{g^2}{f^2} )\mathbf{P}_T
+\frac{1}{\alpha}(P_k({\cal D}_T)+\alpha\frac{g^2}{f^2} )\mathbf{P}_L\right] \\
\end{array} \right)\,.
\ee

To calculate the running of the gauge coupling we can set $D_\mu\varphi^\alpha=0$.
Then $B=0$, the inverse bosonic propagator is diagonal and equation
(\ref{betafunctional}) reduces to
\be
\label{trace1}
\frac{d\Gamma_k}{dt}=
\frac{1}{2}\mathrm{Tr}_\xi\left( \frac{\partial_t{R}_k+\eta_\xi R_k}{P_k+\alpha \frac{g^2}{f^2} } \right)
+\frac{1}{2}\mathrm{Tr}_t\left( \frac{\partial_t{R}_k+\eta_a R_k}{P_k+\frac{g^2}{f^2}  } \right)
+\frac{1}{2}\mathrm{Tr}_\psi\left( \frac{\partial_t{R}_k+\eta_a R_k}{P_k+\alpha\frac{g^2}{f^2}  } \right)
-\mathrm{Tr}_{gh}\left( \frac{\partial_t{R}_k}{P_k+\alpha \frac{g^2}{f^2}  } \right)\, .
\ee
We find
\be
\label{betag2}
\frac{d}{dt}\frac{1}{g^2}=\frac{N}{(4\pi)^{2}} 
\left[\frac{10}{3}\frac{2+\eta_a}{1+\frac{g^2}{\tilde f^2}}+
\frac{1}{3}\left(2-\frac{2+\eta_\xi}{8}\right)\frac{1}{1+\alpha\frac{g^2}{\tilde f^2}}\right]\, .
\ee
The first term is the gauge boson contribution, while in the second term
the first term in bracket is the contribution of the ghosts and 
the second is the contribution of the Goldstone bosons.

The beta function of the Goldstone boson coupling can be computed setting $A_\mu=0$.
Then one can use standard momentum space techniques.
The result is:
\bea
\label{betazeta3}
\frac{d}{dt}\frac{1}{f^2} &=&
\frac{N}{(4\pi)^2}
\Biggl\{
\frac{k^2}{2}\frac{1+\frac{\eta_\xi}{4}}{1+\alpha\frac{g^2}{\tilde f^2} }
+\frac{g^2}{f^2} \Biggl[
\left(1+\frac{\eta_\xi}{6}\right)
\left(\frac{3}{4}\frac{1}{(1+\frac{g^2}{\tilde f^2} )(1+\alpha\frac{g^2}{\tilde f^2} )^2}
+\frac{\alpha}{4}\frac{1}{(1+\alpha\frac{g^2}{\tilde f^2} )^3}\right)
\nonumber\\
&&
\qquad\qquad\qquad\qquad\qquad\ \ \ 
+\left(1+\frac{\eta_a}{6}\right)
\left(\frac{3}{4}\frac{1}{(1+\frac{g^2}{\tilde f^2} )^2(1+\alpha\frac{g^2}{\tilde f^2} )}
+\frac{\alpha}{4}\frac{1}{(1+\alpha\frac{g^2}{\tilde f^2} )^3}\right)
\Biggr]\Biggr\}\,.
\eea

\subsection{Comparison}

We make here some observations concerning the gauge- and scheme-dependence of the beta functions.
Specializing equation (\ref{betag1}) to the case $d=4$ we obtain
\bea
\label{antonio}
\frac{d}{dt}\frac{1}{g^2}&=&
\frac{N}{(4\pi)^2}
\left[\frac{8}{(1+\frac{g^2}{\tilde f^2} )^3}\left(1+\frac{\eta_a}{6}\right)
-\frac{1}{3}\left(\frac{9}{4}+2\eta_a+\frac{1}{8}\eta_\xi\right)\frac{1}{1+\frac{g^2}{\tilde f^2} }
\right]\, . 
\eea
This does not agree with equations (\ref{betag2}),
specialized to the case $\alpha=1$.
However, if we restrict ourselves to the one loop part of the beta function,
{\it i.e.} if we neglect all the terms involving $\eta_\xi$ and $\eta_a$,
and we consider energies much larger than the threshold at $\frac{g^2}{ f^2} $,
then $\frac{g^2}{\tilde f^2} \ll 1$ and
the beta function of the gauge coupling reduces to
\be
\label{simple}
\frac{d}{dt}\frac{1}{g^2}=\frac{N}{(4\pi)^{2}}\frac{29}{4}\ .
\ee
This is the same in both calculations and illustrates
the universality of these beta functions.

On the other hand specializing equation (\ref{betazeta1}) to the case $d=4$ we obtain
\bea
\frac{d}{dt}\frac{1}{f^2}  &=&
\frac{N}{(4\pi)^2}
\frac{k^2}{4}
\frac{1}{(1+\frac{g^2}{\tilde f^2} )^2}
\left[
1+\frac{\eta_\xi}{6} +\frac{4\frac{g^2}{\tilde f^2} }{1+\frac{g^2}{\tilde f^2} }
\left(2+\frac{\eta_\xi+\eta_a}{6}\right)\right] \,.
\eea
At one loop and high energy it becomes
\be
\frac{d}{dt}\frac{1}{ f^2}  =\frac{1}{(4\pi)^{2}}\frac{N}{4}k^2\, ,
\ee
whereas equation (\ref{betazeta3}) reduces in the same limit to
\be
\frac{d}{dt}\frac{1}{f^2}  =\frac{1}{(4\pi)^{2}}\frac{N}{2}k^2\, .
\ee
As already observed in \cite{zanusso}, the difference in the coefficient is the effect of passing from the type I cutoff to a type II cutoff.
Thus, even the leading terms of these beta functions are scheme dependent.
We note however that being the integral of a positive function it is always strictly positive.
Note also that in this approximation, the difference could be absorbed in a redefinition of $k$ if one wanted.
Finally, it is also worth noting that if we assume $g^2/\tilde f^2 \ll 1$,
the beta function of $1/f^2$, in any one of these schemes,
reduces to the one of the pure NL$\sigma$M.


\section{Results}
 

\subsection{Fixed points in $d=4$}

In this subsection we restrict ourselves to $d=4$.
As mentioned before, due to the presence of the terms involving $\eta_a$ and $\eta_\xi$,
equations (\ref{betag1}), (\ref{betazeta1}), (\ref{betag2}), (\ref{betazeta3})
are not the beta functions themselves but linear equations for the beta functions.
They do become the one loop beta functions if one drops all the terms involving
$\eta_a$ and $\eta_\xi$.
Otherwise, before solving for the flow, one has to solve them.
The general structure of the beta functions is
\bea
\frac{d}{dt}\frac{1}{f^2}&=&A_1k^2+B_{11}k^2\eta_\xi+B_{12}k^2\eta_a\ ,
\\
\frac{d}{dt}\frac{1}{g^2}&=&A_2+B_{21}\eta_\xi+B_{22}\eta_a\ ,
\eea
where $A_i$ and $B_{ij}$ are (dimensionless) functions of $\tilde f$ and $g$
that one can easily read off from equations (\ref{betag1}), (\ref{betazeta1}), (\ref{betag2}), 
(\ref{betazeta3}).
The solution of these algebraic equations has the form
\bea
\label{gen1}
\frac{d\tilde f}{dt}&=&
\tilde f-\frac{1}{2}\frac{(A_1+(B_{12}A_2-B_{22}A_1)g^2)\tilde f^3}{1-B_{22}g^2-B_{11}\tilde f^2+(B_{11}B_{22}-B_{12}B_{21})g^2\tilde f^2}
\ ,
\\
\label{gen2}
\frac{dg}{dt}&=&
-\frac{1}{2}\frac{(A_2+(B_{21}A_1-B_{11}A_2)\tilde f^2)g^3}{1-B_{11}\tilde f^2-B_{22}g^2+(B_{11}B_{22}-B_{12}B_{21})\tilde f^2 g^2}
\ .
\eea

Notice that it is the beta functions
of the dimensionless couplings that have to vanish in the definition of fixed point.
In the one loop approximation one just sets all the $B_{ij}$ coefficients
to zero, so that the denominators simplify to one, and in the numerators
only the terms $A_1$ and $A_2$ survive.
Comparison of equations (\ref{betag2}) and (\ref{antonio}) shows that even at one loop
the beta function of $g$ is scheme- and gauge-dependent.
However, this dependence only affects the threshold behavior due to
the fact that this model describes massive gauge fields.
For $k^2\gg g^2/f^2$ the massive modes decouple and this is reflected
in the large denominators, which effectively switch off the beta functions.
If one considers the regime $k^2\gg g^2/f^2$, the denominators
reduce to one.
In this case the beta function of $g$ is given by equation (\ref{simple}):
\be
\frac{dg}{dt}=-\frac{1}{2}A_2 g^3\ ;\qquad 
\ee
with a universal coefficient $A_2=\frac{N}{(4\pi)^2}\frac{29}{4}$.
Note that $29/4$ differs from the coefficient $22/3$ of the pure gauge theory 
by the Goldstone boson contribution $-1/12$.
This contribution is quite small and does not spoil the asymptotic freedom of $g$.
On the other hand, in the same limit the beta function of $\tilde f$ becomes
\be
\frac{d \tilde f}{dt}=\tilde f-\frac{1}{2}A_1 \tilde f^3
\ee
with $A_1=\frac{1}{(4\pi)^2}\frac{N}{4}$ or $A_1=\frac{1}{(4\pi)^2}\frac{N}{2}$
for cutoffs of type I or II respectively.
This beta function has a nontrivial fixed point at $\tilde f_*=\sqrt{2/A_1}$.

The solution of the beta functions (\ref{gen1}) and (\ref{gen2}),
including the ``RG improvement'' (due to the $\eta$-terms) requires a bit more work.
In addition to the Gaussian fixed point at $g=0$, $\tilde f=0$, there is
always also a non-Gaussian fixed point where $\tilde f \not=0$.
The position of this fixed point and the scaling exponents $\theta_i$
(defined as minus the eigenvalues of the linearized flow equations) are given in the following table:

\bigskip

\begin{center}
\begin{tabular}{|c|c|c|c|c|}
\hline
 cutoff and gauge             &$\tilde f_*$ & $g_*$ & $\theta_1$ & $\theta_2$ \\
\hline
type I, $\alpha=1$ & $4\pi\sqrt{6/N}$ & $0$  & $8/3$ & $0$ \\
\hline
type II, $\alpha=1$  & $8\pi\sqrt{2/3N}$ & $0$  & $3$  & $0$  \\
\hline
type II, $\alpha=0$   & $8\pi\sqrt{2/3N}$ & $0$  & $3$  &  $0$ \\
%
\hline
\end{tabular}
\end{center}

\bigskip

This gives an idea of the scheme- dependence of the results.
Note that $g$ is always asymptotically free, and when we set $g=0$
the beta function of $f$ becomes $\alpha$-independent.
Therefore, the position of the fixed point is actually gauge independent.

\subsection{Fixed points in other dimensions}

We briefly consider the solutions of the beta functions 
(\ref{betag1}) and (\ref{betazeta1}) in arbitrary dimension.
The existence of nontrivial fixed points in Yang-Mills theories in $d>4$
has been discussed earlier in \cite{kazakov}.
It is due to the nontrivial dimensionality of the gauge coupling.
One would expect it to be there also in the presence of the Goldstone bosons.
As usual, the simplest way to see this is to consider the one loop beta functions
\bea
\frac{d\tilde f}{dt}&=&\frac{d-2}{2}\tilde f-\frac{1}{2}A_1 \tilde f^3;\qquad 
\\
\frac{d\tilde g}{dt}&=&\frac{d-4}{2}\tilde g-\frac{1}{2}A_2 g^3\ 
\eea
where  $\tilde f=k^{(d-2)/2} f$ and $\tilde g=k^{(d-4)/2}g$ .
From (\ref{betag1}) and (\ref{betazeta1}) one finds
\bea
A_1&=&
\frac{1}{(4\pi)^{d/2}}\frac{N}{2}
\frac{1}{\Gamma\left(\frac{d}{2}+1\right)}
\frac{1}{(1+\frac{\tilde g^2}{\tilde f^2} )^2}
\left[
1+\frac{8{\tilde g^2}/{\tilde f^2}}{1+\frac{\tilde g^2}{\tilde f^2}}
\right] \, ,
\\
A_2&=&
\frac{1}{(4\pi)^{d/2}}\frac{N}{3}
\frac{1}{\Gamma\left(\frac{d}{2}-1\right)}
\frac{1}{1+\frac{\tilde g^2}{\tilde f^2}}
\left[
-\frac{1}{4}-d+2+\frac{192}{d(d-2)}
\frac{1}{(1+\frac{\tilde g^2}{\tilde f^2})^2}\right] 
\eea
In the limit $k^2\gg g^2/f^2$, $A_1$ and $A_2$ become positive constants
implying a fixed point at
\be
\tilde f_*=\sqrt{\frac{d-2}{A_1}}\ ;\qquad
\tilde g_*=\sqrt{\frac{d-4}{A_2}}\ .
\ee
We note that the value of $\frac{\tilde g^2}{\tilde f^2}=\frac{d-4}{d-2}\frac{A_1}{A_2}$ 
at this fixed point
is indeed rather small, so that the approximation is justified a posteriori.
For a better approximation one has to solve the equations numerically.

\subsection{Comments}

The nontrivial fixed point that has been found in these calculations
could be the basis of asymptotic safety in a spontaneously broken chiral theory.
Although its existence was known in the NL$\sigma$M in $2<d<4$ \cite{bardeen},
its presence has been obscured by the widespread use of dimensional regularization.
The use of this regularization method artificially removes power divergences, 
which give important contributions to the beta function of dimensionful couplings such as $f$. These contributions are essential in generating the nontrivial fixed point.
It is enough to use a cutoff regularization at one loop to see the
emergence of the fixed point.
The functional RG techniques used here allow us to go beyond one loop
by resumming infinitely many perturbative contributions.
Further improvements using these techniques can be achieved by
going to higher orders of the derivative expansion.

We have seen here that, within the truncation we work with, the presence of the Goldstone bosons does not affect the asymptotic freedom of the gauge fields,
so that the fixed point in the Goldstone boson sector is the same
as in the ungauged case.
We expect that the same will be true when the four-derivative terms are added.
If this is the case, the fixed point structure of the chiral NL$\sigma$M couplings should be the same as described in \cite{zanusso}.

A somewhat worrying aspect of these results, especially if one restricts oneself
to the one loop approximation, is that they generally require strong interactions.
This follows from the fact that in the beta function of $\tilde f$
the loop contribution has to cancel the classical scaling term.
Addressing this worry is actually the main reason for using functional RG methods: 
their validity does not rely on the coupling being small.
Of course, one is then making other approximations, namely neglecting
higher order terms in the derivative expansion.

If $\tilde f_*<8\pi$ the leading order term of \chipt\ that we have studied is the dominant one,
and due to the existence of the fixed point \chipt\ is convergent at all energies. In EW$\chi$PT, $f$ is related
to the electroweak VEV via the identification $1/f^2=\upsilon^2/4$.
If one follows an RG trajectory towards higher energies one will encounter essentially
two distinct regimes. For energies below the mass of the gauge fields,
the beta functions are suppressed by the threshold terms.
For energies above the mass of the gauge fields the coupling $f$ runs,
behaving asymptotically like $1/k$ and
giving rise to a nearly scale invariant regime (scale invariance is broken by the
running of $g$, which is however very slow in comparison).
The onset of the nearly scale invariant regime depends on the position of the fixed point and occurs earlier for smaller values of $\tilde f_*$.
For instance, if $\tilde f_*=8\pi$, the scale invariant regime
begins at approximately 20 TeV, whereas if $\tilde f_*=2$, the scale invariant regime begins at approximately 1 TeV.



\begin{thebibliography}{99}


\bibitem{ccwz}
S. Weinberg, Phys. Rev. {\bf 166} 1568-1577 (1968);\\
S. Coleman, J. Wess and B. Zumino, Phys. Rev. {\bf 177} 2239 (1969);\\
C. Callan, S. Coleman, J. Wess and B. Zumino, Phys. Rev. {\bf 177} 2247 (1969).\\
A. Salam and J. Strathdee, Phys. Rev. {\bf 184} 1750 (1969).

\bibitem{weinberg1}
S. Weinberg, Physica {\bf A96} 327-340 (1979). 

\bibitem{gl} 
J. Gasser, H. Leutwyler, Annals Phys. {\bf 158} 142 (1984).

\bibitem{ab} 
T. Appelquist, C.W. Bernard, Phys.Rev. {\bf D22} 200, (1980);\\
A.C. Longhitano, Phys. Rev. {\bf D22}, 1166 (1980).
%

\bibitem{chan}
L.H. Chan, Phys. Rev. {\bf D 36} 3755 (1987).
%
\bibitem{herrero}
M.J. Herrero and E. Ruiz Morales, Nucl. Phys. {\bf B 418} 431-455 (1994),
{\it ibid.} {\bf B 437} 319-355 (1995).
%
\bibitem{bagger}
J.A. Bagger, A.F. Falk and M Swartz, 
Phys. Rev. {\bf D76} 105026 (2007).

\bibitem{weinberg2}
 S.~Weinberg, In \emph{General Relativity: An Einstein centenary survey}, ed. S. W. Hawking and W. Israel, pp. 790-831, Cambridge University Press (1979).

\bibitem{gravity}
M. Niedermaier and M. Reuter, Living Rev. Relativity 9, 5  (2006);\\
R. Percacci, 
in ``Approaches to Quantum Gravity: 
Towards a New Understanding of Space, Time and Matter'' 
ed. D. Oriti, Cambridge University Press (2009); 
e-Print: arXiv:0709.3851 [hep-th];\\
D.F. Litim, 
PoS(QG-Ph)024 [ arXiv:0810.3675 [hep-th]];\\
A. Codello, R. Percacci, C. Rahmede,
Annals Phys. {\bf 324} (2009) 414
[arXiv:0805.2909 [hep-th]].


\bibitem{perini}
R.~Percacci and D.~Perini,
Phys.\ Rev.\  D {\bf 68} (2003) 044018 [arXiv:hep-th/0304222];\\
G.P. Vacca and O. Zanusso, Phys. Rev. Lett. {\bf 105} 231601 (2010) arXiv:1009.1735.

\bibitem{gies}
H. Gies and M.M. Scherer, 
Eur. Phys. J. {\bf C66} 387-402 (2010) 
arXiv:0901.2459 [hep-th];\\
H. Gies, S. Rechenberger and M.M. Scherer, 
Eur. Phys. J. {\bf C66} 403-418 (2010) 
arXiv:0907.0327 [hep-th];\\
M.M. Scherer, H. Gies, S. Rechenberger,
Acta Phys. Polon. Proc. Suppl. {\bf 2} 469-694 (2009)
arXiv:0910.0395 [hep-th]

\bibitem{us}
M. Fabbrichesi, R. Percacci, A. Tonero, L. Vecchi and O. Zanusso in preparation.

\bibitem{BG}
J. Honerkamp, Nucl. Phys.  B {\bf 36} 130-140 (1972);\\
L. Alvarez-Gaume, D.Z. Freedman, S. Mukhi, Annals Phys. {\bf 134} 85 (1981);\\
D.G. Boulware and L.S. Brown, Annals Phys. {\bf 138} 392-433 (1982);\\
P.S. Howe, G. Papadopoulos, K.S. Stelle, Nucl. Phys. {\bf B296} 26 (1988).

\bibitem{Wetterich}
C.~Wetterich 
Phys. Lett. B {\bf 301} 90 (1993).

\bibitem{gerge}
  M.~Reuter and C.~Wetterich,
  Nucl.\ Phys.\  B {\bf 417}, 181 (1994).
\\
  F.~Freire, D.~F.~Litim and J.~M.~Pawlowski,
  Phys.\ Lett.\  B {\bf 495} 256 (2000) 
  [arXiv:hep-th/0009110].
  \\
J.M. Pawlowski,
Annals Phys. {\bf 322} 2831-2915 (2007).
e-Print: hep-th/0512261

\bibitem{codello}
A.~Codello and R.~Percacci,
Phys.\ Lett.\  B {\bf 672} 280 (2009)
[arXiv:0810.0715 [hep-th]].
 
\bibitem{zanusso}
R.~Percacci and O.~Zanusso,
Phys. Rev. {\bf D81} 065012 (2010)
arXiv:0910.0851 [hep-th].

\bibitem{cpr2}
A.~Codello , R.~Percacci and C.~Rahmede,
Annals Phys. {\bf 324} 414-469 (2009)
[arXiv:0805.2909 [hep-th]].

\bibitem{optimized}
D.~F.~Litim,
Phys.\ Rev.\  D {\bf 64} (2001) 105007 [arXiv:hep-th/0103195].

\bibitem{kazakov}
D.I. Kazakov
JHEP 03, 020 (2003)
e-Print:hep-th/0209100;\\
H. Gies,
Phys. Rev. D68, 085015 (2003)
e-Print:hep-th/0305208.


\bibitem{bardeen}
  W.~A.~Bardeen, B.~W.~Lee and R.~E.~Shrock,
  Phys.\ Rev.\  D {\bf 14}, 985 (1976).

\end{thebibliography}
\end{document}